\documentclass[10pt]{article}
\usepackage{graphicx}
\usepackage{amsmath}
\usepackage{amssymb}
\usepackage{caption2}
\begin{document}
\begin{center}
{\large {\bf \sc{ $D_{s3}^*(2860)$ and $D_{s1}^*(2860)$ as the 1D $c\bar{s}$ states }}} \\[2mm]
Zhi-Gang Wang  \footnote{E-mail:zgwang@aliyun.com. } \\
  Department of Physics, North China Electric Power University, Baoding 071003, P. R.
  China
\end{center}

\begin{abstract}
In this article, we take the $D_{s3}^*(2860)$ and $D_{s1}^*(2860)$ as the $1^3{\rm D}_3$ and $1^3{\rm D}_1$ $c\bar{s}$ states, respectively,  study   their  strong decays   with the heavy meson  effective theory by including the chiral symmetry breaking corrections. We can reproduce the experimental data  ${\rm Br}\left(D_{sJ}^*(2860)\to D^*K\right)/{\rm Br}\left(D_{sJ}^*(2860)\to D K\right) =1.10 \pm 0.15 \pm 0.19$ with suitable hadronic coupling constants, the assignment of the $D_{sJ}^*(2860)$ as the $D_{s3}^*(2860)$ is favored, the chiral symmetry breaking corrections are large. Furthermore, we obtain  the analytical expressions of the   decay widths, which can be confronted with the experimental data in the future  to fit the unknown coupling constants.
The predictions of the ratios among the   decay widths can be used  to study the decay properties of the $D_{s3}^*(2860)$ and $D_{s1}^*(2860)$ so as to identify them unambiguously.
On the other hand, if the chiral symmetry breaking corrections are small, the large ratio  $R=1.10 \pm 0.15 \pm 0.19$ requires that the $D_{sJ}^*(2860)$ consists of  at least four resonances $D_{s1}^*(2860)$, $D_{s2}^*(2860)$, $D_{s2}^{*\prime}(2860)$,  $D_{s3}^*(2860)$.
\end{abstract}


\section{Introduction}

In 2006, the BaBar  collaboration  observed the $D^*_{sJ}(2860)$ meson with the mass $(2856.6 \pm 1.5  \pm 5.0)\, \rm{ MeV}$ and the width $(48 \pm 7 \pm 10)\, \rm{ MeV}$ in decays  to the final states  $D^0 K^+$ and $D^+K^0_S$ \cite{BaBar2006}. There have been several possible assignments.   Beveren and Rupp assign  the $ D^*_{sJ}(2860)$ to be  the first radial excitation of the $D_{s0}^*(2317)$ based on a coupled-channel model \cite{Beveren2006}.  Colangelo,   Fazio and   Nicotri assign the $ D^*_{sJ}(2860)$ to be  the   $1^3{\rm D}_3$ $c\bar{s}$ state using  the heavy meson effective theory \cite{Colangelo0607}.
Close et al assign the $ D^*_{sJ}(2860)$   to be  the   $2^3{\rm P}_0$   state in a constituent quark model with novel spin-dependent interactions \cite{Close2007}.
Zhang et al assign  the $ D^*_{sJ}(2860)$ to be the $2^3{\rm P}_0$ or $1^3 {\rm D}_3$ state based on the ${}^3{\rm P}_0$ model \cite{Zhang2007};
 Li,   Ma and Liu  share the same interpretation  based on the   Regge phenomenology \cite{Li2007}.
  However, Ebert, Faustov and Galkin observe that the $D_{sJ}^*(2860)$ does
not fit well to the Regge trajectory $D_s^*(2112)$, $D_{s2}^*(2573)$, $D_{sJ}^*(2860)$, $\cdots$ \cite{EFG}.
 Later,  Li and Ma assign the $D^*_{s1}(2700)$ to be the
$1^3{\rm D}_1-2^3{\rm S}_1$ mixing state and the $D^*_{sJ}(2860)$ to be its orthogonal partner, or the $ D^*_{sJ}(2860)$ to be the  $1^3 {\rm D}_3$ state based on the ${}^3{\rm P}_0$ model \cite{Li0911}.
 Zhong and Zhao assign  the $ D^*_{sJ}(2860)$ to be the $1^3{\rm D}_3$ state with some $1^3{\rm D}_2-1^1{\rm D}_2$ mixing component  using the chiral quark model, i.e.
 they assume that the $ D^*_{sJ}(2860)$ arises from two overlapping resonances \cite{Zhong2008,Zhong2010}.
   Vijande,   Valcarce and Fernandez  assign the $ D^*_{sJ}(2860)$ to be the $c\bar{s}-cn\bar{s}\bar{n}$ mixing state \cite{Vijande2009}.
   Chen,   Wang and Zhang  assign the $ D^*_{sJ}(2860)$ to be  the $1^3{\rm D}_3$ state based on
  a semi-classic flux tube model \cite{Chen2009}. Badalian and Bakker assign the $ D^*_{sJ}(2860)$ to be the $1^3{\rm D}_3$ state based on the QCD string model \cite{Badalian2011}.  Guo and Meissner take the $ D^*_{sJ}(2860)$ as the dynamically generated  $D_1(2420)K$ bound state \cite{FKGuo}.

 In 2009, the BaBar  collaboration confirmed the $D^*_{sJ}(2860)$  in the $D^*K$ channel, and measured the ratio $R$ among the branching fractions \cite{BaBar2009},
 \begin{eqnarray}
R&=& \frac{{\rm Br}\left(D_{sJ}^*(2860)\to D^*K\right)}{{\rm Br}\left(D_{sJ}^*(2860)\to D K\right)}=1.10 \pm 0.15 \pm 0.19\, \, .
 \end{eqnarray}
The observation of the  decays $D^*_{sJ}(2860)\to D^*K$ rules out the $J^P=0^+$ assignment \cite{Beveren2006,Close2007,Zhang2007,Li2007}. On the other hand,  if we take the $D^*_{sJ}(2860)$ as the $1^3{\rm D}_3$ state,  Colangelo,   Fazio and   Nicotri obtain the value $R=0.39$ based on the heavy meson effective theory \cite{Colangelo0607},
while in the ${}^3{\rm P}_0$ model, Zhang et al obtain the value $0.59$ \cite{Zhang2007}, Li and Ma obtain the value $0.75$ \cite{Li0911}, Song et al obtain the value $0.55 \sim 0.80$ \cite{Song2014}.
Recently,   Godfrey  and   Jardine obtain the value $0.43$ based on the relativized quark model combined with the pseudoscalar emission decay model \cite{Godfrey2013}.
 The theoretical values differ from    the experimental value greatly.

Recently,  the LHCb collaboration  observed a structure at $2.86\,\rm{GeV}$ with significance of more than 10 standard deviations in the $\overline{D}^0K^-$ mass spectrum in the Dalitz plot analysis of the decays $B_s^0\to \overline{D}^0K^-\pi^+$,  the structure contains both spin-1 and spin-3 components (i.e. the $D_{s1}^{*-}(2860)$ and the $D_{s3}^{*-}(2860)$, respectively), which
  supports an interpretation of these
states being the $J^P =1^-$ and $3^-$ members of the 1D family \cite{LHCb7574,LHCb7712}. The measured masses and widths are
$M_{D_{s3}^*}=(2860.5\pm 2.6 \pm 2.5\pm 6.0)\,\rm{ MeV}$, $M_{D_{s1}^*}=(2859 \pm 12 \pm 6 \pm 23)\,\rm{ MeV}$,
$\Gamma_{D_{s3}^*}=(53 \pm 7 \pm 4 \pm 6)\,\rm{ MeV}$, and $\Gamma_{D_{s1}^*}=(159 \pm 23\pm 27 \pm 72)\,\rm{ MeV}$, respectively.
Furthermore, the LHCb collaboration obtained the conclusion that the $D^*_{sJ}(2860)$ observed by the BaBar collaboration in the inclusive $e^+e^- \to \overline{D}^0K^{-}X$ production  and by the LHCb collaboration in the $pp \to \overline{D}^0K^{-}X$ processes  consists of at least these two resonances \cite{BaBar2009,LHCb1207}.

According to the predictions of the potential models \cite{EFG,GI,PE}, see Table 1, the masses of the 1D $c\bar{s}$ states is about $2.9\,\rm{GeV}$.
      It is reasonable to assign the $D_{s1}^*(2860)$ and   $D_{s3}^*(2860)$  to be the $\rm{1^3D_1}$ and $\rm{1^3D_3}$  $c\bar{s}$ states, respectively \cite{LHCb7574,LHCb7712}.
However,  the theoretical values $R$ differ from    the experimental value greatly in the case of the $D_{s3}^*(2860)$ or the $\rm{1^3D_3}$ assignment of the $D_{sJ}^*(2860)$. In Ref.\cite{Colangelo0607},  Colangelo,   Fazio and   Nicotri take the leading order heavy meson effective Lagrangian. The two-body strong decays $D_{s3}^*(2860) \to D^*K$, $DK$ take place through the relative F-wave,  the final $K$ mesons have the three momenta $p_K=584\,\rm{MeV}$ and $705\,\rm{MeV}$, respectively. The decay widths
\begin{eqnarray}
\Gamma(D_{s3}^*(2860) \to D^*K, DK) &\propto& p_K^7 \, ,
\end{eqnarray}
 where $p_K^7=2.3\times10^{19}\,\rm{MeV}^7$ and $8.6\times10^{19}\,\rm{MeV}^7$ in the decays to the final states  $D^*K$ and $DK$, respectively. Small difference in $p_K$ can lead to large difference in $p_K^7$, so we have to take into account the  heavy quark symmetry breaking corrections and chiral symmetry breaking corrections so as to make robust predictions.

 In this article, we  take into account the  chiral
symmetry-breaking corrections, and study the two-body strong decays of the $D_{s1}^*(2860)$ and $D_{s3}^*(2860)$ with the heavy meson effective  Lagrangian, and try
to reproduce the experimental value $R= 1.10 \pm 0.15 \pm 0.19$ by assigning the $D_{sJ}^{*}(2860)$ to be the $D_{s1}^{*}(2860)$ and the $D_{s3}^{*}(2860)$, respectively.
  Recently, Wu and Huang study the strong decays of the $D_{s0}^*(2317)$ and $D_{s1}^\prime(2460)$ by including the  chiral symmetry breaking corrections \cite{WuHuang}.
 The heavy meson  effective theory have been applied
to identify the charmed  mesons and bottom mesons \cite{Colangelo0607,Wang1009,Wang1308,Colangelo1207,Colangelo0511}, and
to calculate the radiative, vector-meson, two-pion decays
of the heavy quarkonium states \cite{HMET-RV}.

\begin{table}
\begin{center}
\begin{tabular}{|c|c|c|c|c| }\hline\hline
                 & Expt \cite{LHCb7574,LHCb7712} & \cite{EFG}  & \cite{GI}    & \cite{PE}      \\ \hline

$1^3{\rm D}_1$   & 2859                          & 2913        & 2899         & 2913             \\

$1^1{\rm D}_2$   &  --                           & 2931        & 2900         & 2900            \\

$1^3{\rm D}_2$   &  --                           & 2961        & 2926         & 2953             \\

$1^3{\rm D}_3$   & 2860                          & 2871        & 2917         & 2925            \\   \hline\hline
\end{tabular}
\end{center}
\caption{The masses of the 1D $c\bar{s}$ mesons from the potential
models compared to the experimental data.}
\end{table}

The article is arranged as follows:  we derive  the   strong decay widths of the
charmed mesons $D_{s1}^*(2860)$  and $D_{s3}^*(2860)$  with the heavy meson
effective theory in Sect.2; in Sect.3, we present the
 numerical results and discussions; and Sect.4 is reserved for our
conclusions.

\section{ The strong  decays with the heavy meson effective theory }
In the heavy quark limit, the heavy-light  mesons
$Q{\bar q}$  can be  classified in doublets according to the total
angular momentum of the light antiquark ${\vec s}_\ell$,
${\vec s}_\ell= {\vec s}_{\bar q}+{\vec L} $, where the ${\vec
s}_{\bar q}$ and ${\vec L}$ are the spin and orbital angular momentum of the light antiquark, respectively \cite{RevWise}.
In this article, the revelent doublets are the $L=0$ (S-wave) doublet $(P,P^*)$ with $J^P_{s_\ell}=(0^-,1^-)_{\frac{1}{2}}$, and the $L=2$ (D-wave) doublets
      $(P^*_1,P_2)$ and $(P_2,P_3^{ *})$ with
$J^P_{s_\ell}=(1^-,2^-)_{\frac{3}{2}}$ and $(2^-,3^-)_{\frac{5}{2}}$,
respectively.
In the heavy meson effective theory,  those  doublets
can be described by the effective super-fields $H_a$,   $X_a$ and $Y_a$,  respectively \cite{Falk1992},
\begin{eqnarray}
H_a & =& \frac{1+{\rlap{v}/}}{2}\left\{P_{a\mu}^*\gamma^\mu-P_a\gamma_5\right\} \, ,   \nonumber  \\
X_a^\mu &=&\frac{1+{\rlap{v}/}}{2} \Bigg\{ P^{\mu\nu}_{2a}
\gamma_5\gamma_\nu -P^{*}_{1a\nu} \sqrt{3 \over 2} \left[ g^{\mu \nu}-{\gamma^\nu (\gamma^\mu+v^\mu) \over 3}  \right]\Bigg\} \, , \nonumber  \\
Y_a^{ \mu\nu} &=&\frac{1+{\rlap{v}/}}{2} \left\{P^{*\mu\nu\sigma}_{3a} \gamma_\sigma -P^{\alpha\beta}_{2a}\sqrt{5 \over 3} \gamma_5 \left[ g^\mu_\alpha g^\nu_\beta -{g^\nu_\beta\gamma_\alpha  (\gamma^\mu-v^\mu) \over 5} - {g^\mu_\alpha\gamma_\beta  (\gamma^\nu-v^\nu) \over 5}  \right]\right\}\, ,
\end{eqnarray}
where the  heavy meson fields  $P^{(*)}$ contain a factor $\sqrt{M_{P^{(*)}}}$ and
have dimension of mass $\frac{3}{2}$.
The super-fields $H_a$ contain the $\rm{S}$-wave mesons $(P,P^*)$;  $X_a$, $Y_a$ contain  the $\rm{D}$-wave mesons $(P^*_1,P_2)$,
$(P_2,P_3^{ *})$,  respectively.

 The light pseudoscalar mesons are described by the fields
 $\displaystyle \xi=e^{i {\cal M} \over
f_\pi}$, where
\begin{eqnarray}
{\cal M}&=&\lambda^j {{\mathcal{P}}^j}= \left(\begin{array}{ccc}
\sqrt{\frac{1}{2}}\pi^0+\sqrt{\frac{1}{6}}\eta & \pi^+ & K^+\nonumber\\
\pi^- & -\sqrt{\frac{1}{2}}\pi^0+\sqrt{\frac{1}{6}}\eta & K^0\\
K^- & {\bar K}^0 &-\sqrt{\frac{2}{3}}\eta
\end{array}\right) \, ,
\end{eqnarray}
and the decay constant $f_\pi=130\,\rm{MeV}$.

At the leading order approximation, the heavy meson chiral Lagrangians   ${\cal
L}_{X}$ and ${\cal L}_{Y}$  for
the strong decays to the light pseudoscalar mesons  can be  written as:
\begin{eqnarray}
{\cal L}_{X} &=& {g_{X} \over \Lambda}{\rm Tr}\left\{{\bar H}_a X^\mu_b(i {\cal D}_\mu {\not\! {\cal A}  }+i{\not\! {\cal D}  } { \cal A}_\mu)_{ba} \gamma_5\right\}  +h.c.  \, , \nonumber\\
{\cal L}_{Y} &=&  {1 \over {\Lambda^2}}{\rm Tr}\left\{ {\bar H}_a Y^{\mu \nu}_b \left[g_{Y} \{i{\cal D}_\mu, i{\cal D}_\nu\} {\cal A}_\lambda + \tilde{g}_{Y} \left(i{\cal D}_\mu i{\cal D}_\lambda { \cal A}_\nu + i{\cal D}_\nu i{\cal D}_\lambda { \cal A}_\mu \right)\right]_{ba}  \gamma^\lambda \gamma_5\right\}+h.c.\, ,
\end{eqnarray}
where
\begin{eqnarray}
{\cal D}_{\mu}&=&\partial_\mu+{\cal V}_{\mu} \, , \nonumber \\
 {\cal V}_{\mu }&=&\frac{1}{2}\left(\xi^\dagger\partial_\mu \xi+\xi\partial_\mu \xi^\dagger\right)\, , \nonumber \\
 {\cal A}_{\mu }&=&\frac{1}{2}\left(\xi^\dagger\partial_\mu \xi-\xi\partial_\mu  \xi^\dagger\right)\,  , \nonumber\\
 \{ {\cal D}_\mu, {\cal D}_\nu \}&=&{\cal D}_\mu {\cal D}_\nu+{\cal D}_\nu {\cal D}_\mu  \, ,
 \end{eqnarray}
the hadronic coupling constants $g_{X}$, $g_{Y}$ and $\tilde{g}_{Y}$ are parameters and   can be fitted  to the experimental data \cite{HL,PRT1997}, $\Lambda$ is the chiral symmetry breaking scale and chosen as
$\Lambda = 1 \, \rm{GeV} $  \cite{Colangelo0511}.

We construct the chiral symmetry breaking
  Lagrangians ${\cal L}_{X}^\chi$ and ${\cal L}_{Y}^\chi$    according   to Refs.\cite{Heavy-Chiral,Heavy-Chiral-2},
\begin{eqnarray}
{\cal L}^\chi_{X}&=& {k^1_{X} \over \Lambda^2}{\rm Tr}\left\{{\bar H}_a X^\mu_b(i {\cal D}_\mu {\not\! {\cal A}  }+i{\not\! {\cal D}  } { \cal A}_\mu)_{bc}
\left(m_q^\xi \right)_{ca}\gamma_5\right\}   \nonumber\\
&&+ {k^2_{X} \over \Lambda^2}{\rm Tr}\left\{{\bar H}_a X^\mu_c\left(m_q^\xi \right)_{cb}(i {\cal D}_\mu {\not\! {\cal A}  }+i{\not\! {\cal D}  } { \cal A}_\mu)_{ba}\gamma_5\right\}   \nonumber   \\
&&+ {k^3_{X} \over \Lambda^2}{\rm Tr}\left\{{\bar H}_a X^\mu_b(i {\cal D}_\mu {\not\! {\cal A}  }+i{\not\! {\cal D}  } { \cal A}_\mu)_{ba}
\left(m_q^\xi \right)_{cc}\gamma_5\right\}   \nonumber   \\
&&+ {k^4_{X} \over \Lambda^2}{\rm Tr}\left\{{\bar H}_a X^\mu_a(i {\cal D}_\mu {\not\! {\cal A}  }+i{\not\! {\cal D}  } { \cal A}_\mu)_{bc}\left(m_q^\xi \right)_{cb}\gamma_5\right\}   \nonumber   \\
&&+ {1 \over {\Lambda^2}}{\rm Tr}\left\{ {\bar H}_a X^{\mu }_b   \left[k^5_{X} \{i{\cal D}_\mu, iv\cdot{\cal D} \} {\cal A}_\lambda + \tilde{k}_{X}^5  \left\{iv\cdot{\cal D} , i{\cal D}_\lambda \right\}{ \cal A}_\mu +\tilde{ \tilde{k}}_{X}^5  \left\{ i{\cal D}_\mu , i{\cal D}_\lambda \right\}{v\cdot \cal A}   \right]_{ba}  \gamma^\lambda \gamma_5\right\}\nonumber\\
&&+h.c. \, \, ,
\end{eqnarray}

\begin{eqnarray}
{\cal L}_{Y}^\chi&=&  {1 \over {\Lambda^3}}{\rm Tr}\left\{ {\bar H}_a Y^{\mu \nu}_b \left[k^1_{Y} \{i{\cal D}_\mu, i{\cal D}_\nu\} {\cal A}_\lambda + \tilde{k}^1_{Y} \left(i{\cal D}_\mu i{\cal D}_\lambda { \cal A}_\nu + i{\cal D}_\nu i{\cal D}_\lambda { \cal A}_\mu \right)\right]_{bc} \left(m_q^\xi\right)_{ca} \gamma^\lambda \gamma_5\right\}  \nonumber   \\
&&+  {1 \over {\Lambda^3}}{\rm Tr}\left\{ {\bar H}_a Y^{\mu \nu}_b \left(m_q^\xi\right)_{bc}\left[k^2_{Y} \{i{\cal D}_\mu, i{\cal D}_\nu\} {\cal A}_\lambda + \tilde{k}^2_{Y} \left(i{\cal D}_\mu i{\cal D}_\lambda { \cal A}_\nu + i{\cal D}_\nu i{\cal D}_\lambda { \cal A}_\mu \right)\right]_{ca}  \gamma^\lambda \gamma_5\right\}  \nonumber   \\
&&+  {1 \over {\Lambda^3}}{\rm Tr}\left\{ {\bar H}_a Y^{\mu \nu}_b \left[k^3_{Y} \{i{\cal D}_\mu, i{\cal D}_\nu\} {\cal A}_\lambda + \tilde{k}^3_{Y} \left(i{\cal D}_\mu i{\cal D}_\lambda { \cal A}_\nu + i{\cal D}_\nu i{\cal D}_\lambda { \cal A}_\mu \right)\right]_{ba} \left(m_q^\xi\right)_{cc} \gamma^\lambda \gamma_5\right\}  \nonumber   \\
&&+  {1 \over {\Lambda^3}}{\rm Tr}\left\{ {\bar H}_a Y^{\mu \nu}_a \left[k^4_{Y} \{i{\cal D}_\mu, i{\cal D}_\nu\} {\cal A}_\lambda + \tilde{k}^4_{Y} \left(i{\cal D}_\mu i{\cal D}_\lambda { \cal A}_\nu + i{\cal D}_\nu i{\cal D}_\lambda { \cal A}_\mu \right)\right]_{bc} \left(m_q^\xi\right)_{cb} \gamma^\lambda \gamma_5\right\}  \nonumber   \\
&&+{1 \over {\Lambda^3}}{\rm Tr}\left\{ {\bar H}_a Y^{\mu \nu  }_b \left[k^5_Y  \{i{\cal D}_\mu, i{\cal D}_\nu,i v\cdot{\cal D}  \} {\cal A}_\lambda + \tilde{k}^5_Y \left(\{i{\cal D}_\mu,iv\cdot{\cal D} \} i{\cal D}_\lambda { \cal A}_\nu + \{i{\cal D}_\nu,iv\cdot{\cal D} \} i{\cal D}_\lambda { \cal A}_\mu     \right.\right.\right.\nonumber\\
&&\left.\left.\left.+ \{i{\cal D}_\mu,i{\cal D}_\nu\} i{\cal D}_\lambda v\cdot{ \cal A} \right)\right]_{ba}  \gamma^\lambda \gamma_5\right\} + h.c. \, \, ,
  \end{eqnarray}
where
\begin{eqnarray}
  \{ {\cal D}_\mu, {\cal D}_\nu, {\cal D}_\rho \}&=&{\cal D}_\mu {\cal D}_\nu{\cal D}_\rho+{\cal D}_\mu {\cal D}_\rho{\cal D}_\nu+{\cal D}_\nu {\cal D}_\mu{\cal D}_\rho +{\cal D}_\nu {\cal D}_\rho{\cal D}_\mu+{\cal D}_\rho{\cal D}_\mu {\cal D}_\nu+{\cal D}_\rho{\cal D}_\nu {\cal D}_\mu \, ,
\end{eqnarray}
  $m_q={\rm diag}(m_u,m_d,m_s)$, $m_q^\xi=\xi m_q \xi+\xi^\dagger m_q \xi^\dagger$, $v^\mu=(1,0,0,0)$,  the hadronic coupling constants   $k^j_{X/Y}$, $\tilde{k}^j_{Y}$, $\tilde{k}^{5}_{X}$, $\tilde{\tilde{k}}^{5}_{X}$ with $j=1,5$ can be fitted  to the experimental data.  The flavor and spin violation corrections of the order
$\mathcal {O}(1/m_Q)$ are neglected, as there are too many unknown couplings to be determined,   we expect that the corrections are not    as large as the chiral symmetry breaking corrections. At the hadronic level, the $1/m_Q$ corrections can be  crudely  estimated  to be  of the order $p_K/M_{D^*_{sJ}}\approx 0.20-0.25$ or $\left(M_{D^*_{sJ}}-\overline{M}_{D_s}\right)/M_{D^*_{sJ}}\approx 0.27$ with $\overline{M}_{D_s}=\left(3M_{D^*_s}+M_{D_s} \right)/4$. We can also take into account the  $1/m_Q$ corrections consistently by resorting to the covariant heavy meson chiral theory \cite{LSGeng},  however, we have no experimental  data or
lattice QCD data to fit the unknown hadronic constants, and it is beyond the present work.

From the heavy meson chiral Lagrangians  ${\cal L}_{X}$, ${\cal L}_{Y}$, ${\cal L}_{X}^\chi$ and ${\cal L}_{Y}^\chi$, we can obtain the   decay widths
$\Gamma$ of the   strong decays to the light pseudoscalar mesons,

$\bullet$  $(1^-,2^-)_{\frac{3}{2}}\to (0^-,1^-)_{\frac{1}{2}}+ \mathcal{P}_j$,
\begin{eqnarray}
\Gamma(2^- \to 1^-+\mathcal{P}_j) &=&  \frac{ M_f\left(p_f^2+m_{\mathcal{P}_j}^2\right) p_f^3}{6\pi   M_i } F_j^2\, , \\
\Gamma(1^- \to 1^-+\mathcal{P}_j) &=&  \frac{ M_f\left(p_f^2+m_{\mathcal{P}_j}^2\right) p_f^3}{18\pi   M_i }F_j^2 \, , \\
\Gamma(1^- \to 0^-+\mathcal{P}_j) &=&  \frac{ M_f\left(p_f^2+m_{\mathcal{P}_j}^2\right) p_f^3}{9\pi   M_i }F_j^2 \, ,
\end{eqnarray}
where
\begin{eqnarray}
F_j&=&\frac{2g_X}{f_\pi\Lambda}\lambda^j_{ba}+\frac{4k^1_X}{f_\pi\Lambda^2}\lambda^j_{bc}(m_q)_{ca}+\frac{4k^2_X}{f_\pi\Lambda^2}(m_q)_{bc}\lambda^j_{ca}
+\frac{4k^3_X}{f_\pi\Lambda^2}\lambda^j_{ba}(m_q)_{cc}+\frac{4k^4_X}{f_\pi\Lambda^2}\delta_{ba}\lambda^j_{cd}(m_q)_{dc}\nonumber\\
&&-\frac{2(k^5_X+\tilde{k}^5_X+\tilde{\tilde{k}}^5_X)\sqrt{p_f^2+m_{\mathcal{P}_j}^2}}{f_\pi\Lambda^2}\lambda^j_{ba}\, \, ,\nonumber\\
p_f&=&\frac{\sqrt{(M_i^2-(M_f+m_{\mathcal{P}_j})^2)(M_i^2-(M_f-m_{\mathcal{P}_j})^2)}}{2M_i}\, ,
\end{eqnarray}
 the $i$ (or $b$) and $f$ (or $a$) denote the initial and final state heavy mesons, respectively.

$\bullet$  $(2^-,3^-)_{\frac{5}{2}}\to (0^-,1^-)_{\frac{1}{2}}+ \mathcal{P}_j$,
\begin{eqnarray}
\Gamma(3^- \to 1^-+\mathcal{P}_j) &=&  \frac{4M_f p_f^7}{105\pi   M_i}F_j^2\, , \\
\Gamma(3^- \to 0^-+\mathcal{P}_j) &=&  \frac{ M_f p_f^7}{35\pi   M_i}F_j^2 \, , \\
\Gamma(2^- \to 1^-+\mathcal{P}_j) &=&  \frac{ M_f p_f^7}{15\pi   M_i}F_j^2 \, ,
\end{eqnarray}
where
\begin{eqnarray}
F_j&=&\frac{2(g_Y+\tilde{g}_Y)}{f_\pi\Lambda^2}\lambda^j_{ba}+\frac{4(k^1_Y+\tilde{k}^1_Y)}{f_\pi\Lambda^3}\lambda^j_{bc}(m_q)_{ca}
+\frac{4(k^2_Y+\tilde{k}^2_Y)}{f_\pi\Lambda^3}(m_q)_{bc}\lambda^j_{ca}+\frac{4(k^3_Y+\tilde{k}^3_Y)}{f_\pi\Lambda^3}\lambda^j_{ba}(m_q)_{cc}
\nonumber\\
&&+\frac{4(k^4_Y+\tilde{k}^4_Y)}{f_\pi\Lambda^3}\delta_{ba}\lambda^j_{cd}(m_q)_{dc}-\frac{6(k^5_Y
+\tilde{k}^5_Y)\sqrt{p_f^2+m_{\mathcal{P}_j}^2}}{f_\pi\Lambda^3}\lambda^j_{ba}\, \, .
\end{eqnarray}
In those decays, the energy release $E_{\mathcal{P}_j}=\sqrt{p_f^2+m_{\mathcal{P}_j}^2}$ is rather large, the
chiral expansion does  not converge very quickly, the next-to-leading order chiral  corrections maybe manifest themselves.

\section{Numerical Results and Discussions}
The input parameters are taken  as
$M_{K^+}=493.677\,\rm{MeV}$, $M_{K^0}=497.614\,\rm{MeV}$, $M_{\eta}=547.862\,\rm{MeV}$,
$M_{D^+}=1869.5\,\rm{MeV}$, $M_{D^0}=1864.91\,\rm{MeV}$,
$M_{D_s^+}=1969\,\rm{MeV}$, $M_{D^{*+}}=2010.29\,\rm{MeV}$,
$M_{D^{*0}}=2006.99\,\rm{MeV}$, $M_{D_s^{*+}}=2112.3\,\rm{MeV}$ from the Particle Data Group \cite{PDG}.

We redefine the hadronic coupling constants  $\bar{g}_Y=g_Y+\tilde{g}_Y$, $\bar{k}_Y^j=\left(k_Y^j+\tilde{k}_Y^j\right)/\bar{g}_Y$, $j=1-5$, and write down the decay widths of the $D_{s3}^*(2860)$ explicitly from Eqs.(13-14).
\begin{eqnarray}
\Gamma(D_{s3}^{*+} \to D^{*0}+K^+) &=&  \frac{16\bar{g}_{Y}^2M_{D^*} p_K^7}{105\pi f_{\pi}^2\Lambda^4 M_{D_{s3}^{*}}}\left( 1+\frac{2m_u\bar{k}^1_Y}{\Lambda}+\frac{2m_s\bar{k}^2_Y}{\Lambda}+\frac{2(m_u+m_d+m_s)\bar{k}^3_Y}{\Lambda}\right.\nonumber\\
 &&\left.-\frac{3 \sqrt{p_K^2+m_K^2}\bar{k}^5_Y}{\Lambda}\right)^2\, , \nonumber\\
 \Gamma(D_{s3}^{*+} \to D^{0}+K^+) &=&  \frac{4\bar{g}_{Y}^2M_{D} p_K^7}{35\pi f_{\pi}^2\Lambda^4 M_{D_{s3}^{*}} } \left( 1+\frac{2m_u\bar{k}^1_Y}{\Lambda}+\frac{2m_s\bar{k}^2_Y}{\Lambda}+\frac{2(m_u+m_d+m_s)\bar{k}^3_Y}{\Lambda}\right.\nonumber\\
 &&\left.-\frac{3 \sqrt{p_K^2+m_K^2}\bar{k}^5_Y}{\Lambda}\right)^2\, ,
 \end{eqnarray}

 \begin{eqnarray}
\Gamma(D_{s3}^{*+} \to D^{*+}+K^0) &=&  \frac{16\bar{g}_{Y}^2M_{D^*} p_K^7}{105\pi f_{\pi}^2\Lambda^4 M_{D_{s3}^{*}}}\left( 1+\frac{2m_d\bar{k}^1_Y}{\Lambda}+\frac{2m_s\bar{k}^2_Y}{\Lambda}+\frac{2(m_u+m_d+m_s)\bar{k}^3_Y}{\Lambda}\right.\nonumber\\
 &&\left.-\frac{3 \sqrt{p_K^2+m_K^2}\bar{k}^5_Y}{\Lambda}\right)^2\, ,\nonumber \\
 \Gamma(D_{s3}^{*+} \to D^{+}+K^0) &=&  \frac{4\bar{g}_{Y}^2M_{D} p_K^7}{35\pi f_{\pi}^2\Lambda^4 M_{D_{s3}^{*}} } \left( 1+\frac{2m_d\bar{k}^1_Y}{\Lambda}+\frac{2m_s\bar{k}^2_Y}{\Lambda}+\frac{2(m_u+m_d+m_s)\bar{k}^3_Y}{\Lambda}\right.\nonumber\\
 &&\left.-\frac{3 \sqrt{p_K^2+m_K^2}\bar{k}^5_Y}{\Lambda}\right)^2\, ,
 \end{eqnarray}

\begin{eqnarray}
\Gamma(D_{s3}^{*+} \to D^{*+}_s+\eta) &=&  \frac{32\bar{g}_{Y}^2M_{D_s^*} p_\eta^7}{315\pi f_{\pi}^2\Lambda^4 M_{D_{s3}^*}}\left( 1+\frac{2m_s\bar{k}^1_Y}{\Lambda}+\frac{2m_s\bar{k}^2_Y}{\Lambda}+\frac{2(m_u+m_d+m_s)\bar{k}^3_Y}{\Lambda}\right.\nonumber\\
 &&\left.-\frac{(m_u+m_d-2m_s)\bar{k}^4_Y}{\Lambda}-\frac{3 \sqrt{p_\eta^2+m_\eta^2}\bar{k}^5_Y}{\Lambda}\right)^2\, , \nonumber  \\
 \Gamma(D_{s3}^{*+} \to D^{+}_s+\eta) &=&  \frac{8\bar{g}_{Y}^2M_{D} p_\eta^7}{105\pi f_{\pi}^2\Lambda^4 M_{D_{s3}^*}}\left( 1+\frac{2m_s\bar{k}^1_Y}{\Lambda}+\frac{2m_s\bar{k}^2_Y}{\Lambda}+\frac{2(m_u+m_d+m_s)\bar{k}^3_Y}{\Lambda}\right.\nonumber\\
 &&\left.-\frac{(m_u+m_d-2m_s)\bar{k}^4_Y}{\Lambda}-\frac{3 \sqrt{p_\eta^2+m_\eta^2}\bar{k}^5_Y}{\Lambda}\right)^2\, .
\end{eqnarray}
In this article, we neglect the one-loop chiral corrections, which are estimated to be of the order $\frac{m^2_{\mathcal{P}_j}}{16\pi^2 f_\pi^2}< 10\%$. In the case of the hadronic coupling constants $g$ and $h$ in the heavy meson chiral theory, the one-loop chiral corrections are less than $10\%$ \cite{Heavy-Chiral-2}, which are consistent with the crude estimation.

We define the ratios  $R_{0+}$, $R_{+0}$, $R_s$ among the   decay widths,
 \begin{eqnarray}
 R_{0+}&=&\frac{\Gamma(D_{s3}^{*+} \to D^{*0}+K^+)}{\Gamma(D_{s3}^{*+} \to D^{0}+K^+)} \, , \nonumber\\
 R_{+0}&=&\frac{\Gamma(D_{s3}^{*+} \to D^{*+}+K^0)}{\Gamma(D_{s3}^{*+} \to D^{+}+K^0)} \, , \nonumber\\
 R_{s}&=&\frac{\Gamma(D_{s3}^{*+} \to D^{*+}_s+\eta)}{\Gamma(D_{s3}^{*+} \to D^{+}_s+\eta)} \, .
 \end{eqnarray}
The ratios  $R_{0+}$, $R_{+0}$, $R_s$  are independent on the hadronic coupling constants $\bar{g}_Y$, $\bar{k}^1_Y$, $\bar{k}^2_Y$, $\bar{k}^3_Y$, $\bar{k}^4_Y$,  we can absorb the coupling constants $\bar{k}^1_Y$, $\bar{k}^2_Y$, $\bar{k}^3_Y$, $\bar{k}^4_Y$   into the effective coupling $\bar{g}_Y$ or set  $\bar{k}^1_Y=\bar{k}^2_Y=\bar{k}^3_Y=\bar{k}^4_Y=0$.

Firstly, let us assign the $D_{sJ}^*(2860)$ to be the $D_{s3}^*(2860)$, then we can  obtain the value,
\begin{eqnarray}
\bar{k}^5_Y&=&0.33223\pm0.01248\, ,
 \end{eqnarray}
 by setting the $\frac{R_{0+}+R_{+0}}{2}$ to be the experimental data,  $\frac{R_{0+}+R_{+0}}{2}=R=1.10 \pm 0.15 \pm 0.19$ \cite{BaBar2009}.
On the other hand, if we retain only the leading order coupling constant $\bar{g}_Y$, then $\frac{R_{0+}+R_{+0}}{2}=0.3866$, which is consistent with the value $0.39$ obtained by  Colangelo,   Fazio and   Nicotri   \cite{Colangelo0607}.
The value of the hadronic coupling constant $k_Y^5+\tilde{k}_Y^5$ in the chiral symmetry breaking
  Lagrangian  is about $\frac{1}{3}$ of that of the hadronic coupling constant $g_Y+\tilde{g}_Y$ in the leading-order
  Lagrangian according to the relation   $\bar{k}_Y^5=(k_Y^5+\tilde{k}_Y^5)/(g_Y+\tilde{g}_Y)$. However, taking into account
  such  chiral symmetry breaking term  can enlarge the ratio $R$ about $2.8$ times.

Then we write down the prediction of the ratio $R_s$,
\begin{eqnarray}
R_{s}&=& 0.42 \pm0.06\,\, (0.18) \, ,
\end{eqnarray}
the value $0.18$ in the bracket comes  from the leading order heavy meson effective Lagrangian ${\cal{L}}_Y$, i.e. only the $\bar{g}_Y$ is retained.
The chiral symmetry breaking corrections are rather large, the present predictions  can be confronted with the experimental data in the futures to study the chiral symmetry breaking corrections.

We can also define  the ratios $R_{+}^0$  and $\widetilde{R}_{+}^0$,
\begin{eqnarray}
R_{+}^0&=&\frac{\Gamma(D_{s3}^{*+} \to D^{*0}+K^+)}{\Gamma(D_{s3}^{*+} \to D^{*+}+K^0)}\, , \nonumber \\
\widetilde{R}_{+}^0&=&\frac{\Gamma(D_{s3}^{*+} \to D^{0}+K^+)}{\Gamma(D_{s3}^{*+} \to D^{+}+K^0)}\, ,
\end{eqnarray}
which are sensitive to the chiral symmetry breaking corrections associate with $\bar{k}_Y^1$. We can estimate the $\bar{k}_Y^1$ by   confronting the ratios $R_{+}^0$  and $\widetilde{R}_{+}^0$ with  the experimental data in the future.

Now we assign the $D_{sJ}^*(2860)$ to be the $D_{s1}^*(2860)$ and study the   strong decays of the $D_{s1}^*(2860)$ as the $1^3\rm{D}_1$ $c\bar{s}$ state. Firstly, let us redefine the hadronic coupling constants   $\bar{k}_X^j=k_X^j/g_X$, $j=1-4$, $\bar{k}_X^5=(k_X^5+\tilde{k}_X^5+\tilde{\tilde{k}}_X^5)/g_X$, and write down the decay widths explicitly from Eqs.(10-11),
\begin{eqnarray}
\Gamma(D_{s1}^{*+} \to D^{*0}+K^+) &=&  \frac{2 g_{X}^2M_{D^*}\left(p_K^2+m_K^2\right) p_K^3}{9\pi f_{\pi}^2\Lambda^2 M_{D_{s1}^*} } \left( 1+\frac{2m_u\bar{k}^1_X}{\Lambda}+\frac{2m_s\bar{k}^2_X}{\Lambda}+\frac{2(m_u+m_d+m_s)\bar{k}^3_X}{\Lambda}\right.\nonumber\\
 &&\left.-\frac{ \sqrt{p_K^2+m_K^2}\bar{k}^5_X}{\Lambda}\right)^2\, ,\nonumber \\
\Gamma(D_{s1}^{*+} \to D^{0}+K^+) &=& \frac{4g_{X}^2M_{D}\left(p_K^2+m_K^2\right) p_K^3}{9\pi f_{\pi}^2\Lambda^2 M_{D_{s1}^*} } \left( 1+\frac{2m_u\bar{k}^1_X}{\Lambda}+\frac{2m_s\bar{k}^2_X}{\Lambda}+\frac{2(m_u+m_d+m_s)\bar{k}^3_X}{\Lambda}\right.\nonumber\\
 &&\left.-\frac{ \sqrt{p_K^2+m_K^2}\bar{k}^5_X}{\Lambda}\right)^2\, ,
\end{eqnarray}

\begin{eqnarray}
\Gamma(D_{s1}^{*+} \to D^{*+}+K^0) &=&  \frac{2 g_{X}^2M_{D^*}\left(p_K^2+m_K^2\right) p_K^3}{9\pi f_{\pi}^2\Lambda^2 M_{D_{s1}^*} } \left( 1+\frac{2m_d\bar{k}^1_X}{\Lambda}+\frac{2m_s\bar{k}^2_X}{\Lambda}+\frac{2(m_u+m_d+m_s)\bar{k}^3_X}{\Lambda}\right.\nonumber\\
 &&\left.-\frac{ \sqrt{p_K^2+m_K^2}\bar{k}^5_X}{\Lambda}\right)^2\, ,\nonumber \\
\Gamma(D_{s1}^{*+} \to D^{+}+K^0) &=& \frac{4g_{X}^2M_{D}\left(p_K^2+m_K^2\right) p_K^3}{9\pi f_{\pi}^2\Lambda^2 M_{D_{s1}^*} } \left( 1+\frac{2m_d\bar{k}^1_X}{\Lambda}+\frac{2m_s\bar{k}^2_X}{\Lambda}+\frac{2(m_u+m_d+m_s)\bar{k}^3_X}{\Lambda}\right.\nonumber\\
 &&\left.-\frac{ \sqrt{p_K^2+m_K^2}\bar{k}^5_X}{\Lambda}\right)^2\, ,
\end{eqnarray}

\begin{eqnarray}
\Gamma(D_{s1}^{*+} \to D^{*+}_s+\eta) &=&  \frac{4 g_{X}^2M_{D_s^*}\left(p_\eta^2+m_\eta^2\right) p_\eta^3}{27\pi f_{\pi}^2\Lambda^2 M_{D_{s1}^*} } \left( 1+\frac{2m_s\bar{k}^1_X}{\Lambda}+\frac{2m_s\bar{k}^2_X}{\Lambda}+\frac{2(m_u+m_d+m_s)\bar{k}^3_X}{\Lambda}\right.\nonumber\\
 &&\left.-\frac{(m_u+m_d-2m_s)\bar{k}^4_X}{\Lambda}-\frac{ \sqrt{p_\eta^2+m_\eta^2}\bar{k}^5_X}{\Lambda}\right)^2\, , \nonumber\\
\Gamma(D_{s1}^{*+} \to D^{+}_s+\eta) &=& \frac{8g_{X}^2M_{D_s}\left(p_\eta^2+m_\eta^2\right) p_\eta^3}{27\pi f_{\pi}^2\Lambda^2 M_{D_{s1}^*} } \left( 1+\frac{2m_s\bar{k}^1_X}{\Lambda}+\frac{2m_s\bar{k}^2_X}{\Lambda}+\frac{2(m_u+m_d+m_s)\bar{k}^3_X}{\Lambda}\right.\nonumber\\
 &&\left.-\frac{(m_u+m_d-2m_s)\bar{k}^4_X}{\Lambda}-\frac{ \sqrt{p_\eta^2+m_\eta^2}\bar{k}^5_X}{\Lambda}\right)^2\, .
\end{eqnarray}

Then we define the ratio $\overline{R}$,
\begin{eqnarray}
\overline{R}&=& \frac{1}{2}\left\{\frac{\Gamma(D_{s1}^{*+} \to D^{*0}+K^+)}{\Gamma(D_{s1}^{*+} \to D^{0}+K^+)}+\frac{\Gamma(D_{s1}^{*+} \to D^{*+}+K^0)}{\Gamma(D_{s1}^{*+} \to D^{+}+K^0)} \right\} \, ,
\end{eqnarray}
which is independent on the hadronic coupling constants $g_X$, $\bar{k}^1_X$, $\bar{k}^2_X$, $\bar{k}^3_X$, $\bar{k}^4_X$.
We can also absorb the coupling constants $\bar{k}^1_X$, $\bar{k}^2_X$, $\bar{k}^3_X$, $\bar{k}^4_X$   into the effective coupling $g_X$ or set  $\bar{k}^1_X=\bar{k}^2_X=\bar{k}^3_X=\bar{k}^4_X=0$. By setting $\overline{R}=R=1.10 \pm 0.15 \pm 0.19$ \cite{BaBar2009},
 we can  obtain the value,
\begin{eqnarray}
\bar{k}^5_X&=&1.0555\pm0.01953\, .
 \end{eqnarray}
The value of the hadronic coupling constant $k_X^5+\tilde{k}_X^5+\tilde{\tilde{k}}_X^5$ in the chiral symmetry breaking
  Lagrangian  is as large as  that of the hadronic coupling constant $g_X$ in the leading-order
  Lagrangian according to the relation   $\bar{k}_X^5=(k_X^5+\tilde{k}_X^5+\tilde{\tilde{k}}_X^5)/g_X$. The dimensionless coupling constants $\bar{k}^5_X$ and $\bar{k}^5_Y$ are normalized in the same way, and $\bar{k}^5_X\gg\bar{k}^5_Y$, the convergent behavior is much better in the chiral expansion  in case of assigning     the $D_{sJ}^*(2860)$ to be  the $D_{s3}^*(2860)$. The assignment of the $D_{sJ}^*(2860)$ as the $D_{s1}^*(2860)$ is not favored but not excluded, because  a larger   coupling constant $\bar{k}^5_X$ does not mean that  the chiral expansion breaks  down.

Those strong  decays of the  $D_{s1}^*(2860)$ take place through the relative P-wave, the decay widths are  proportional to $p_f^3$, while the strong decays of the $D_{s3}^*(2860)$ take place through the relative F-wave, the decay widths are  proportional to $p_f^7$. The   decay widths of the $D_{s1}^*(2860)$ are much  insensitive to the $p_f$ compared to that of the $D_{s3}^*(2860)$. At the present time, there is no experimental data to fit the hadronic coupling constants.  In the leading order approximation, i.e. we neglect the chiral symmetry breaking  corrections, the ratios $\overline{R}$, $\widetilde{R}$ and $ \overline{R}_{s}$ among the decay widths are
\begin{eqnarray}
\overline{R}&=&0.24\,\,\,(0.46 \sim 0.70)\, ,
\end{eqnarray}
\begin{eqnarray}
\widetilde{ R}&=& \frac{\Gamma(D_{s1}^{*+} \to D^{+}_s+\eta)}{\Gamma(D_{s1}^{*+} \to D^{0}+K^+)+\Gamma(D_{s1}^{*+} \to D^{+}+K^0)}=0.177\,\,\,(0.10 \sim 0.14) \, ,
\end{eqnarray}
 \begin{eqnarray}
\overline{R}_{s}&=&\frac{\Gamma(D_{s1}^{*+} \to D^{*+}_s+\eta)}{\Gamma(D_{s1}^{*+} \to D^{+}_s+\eta)}=0.17 \, ,
\end{eqnarray}
where in the bracket we present the values from the recent studies based on  the ${}^3{\rm P}_0$ model \cite{Song2014}. Also in the ${}^3{\rm P}_0$ model,   Zhang et al obtain the value $\overline{R}=0.16$ \cite{Zhang2007}. The present value $\overline{R}$ differs greatly from that obtained in Ref.\cite{Song2014}, while it is compatible with that obtained in Ref.\cite{Zhang2007}. In Ref.\cite{Godfrey2013}, Godfrey  and   Jardine obtain the value $0.34$ based on the relativized quark model combined with  the pseudoscalar emission decay model, which is larger than the present calculation. The present predictions can be confronted with the experimental data in the future to study the strong decays  of the $D_{s1}^*(2860)$.

In the leading order approximation, we obtain the values $R=0.39$ and $\overline{R}=0.24$ in the cases  of assigning the $D_{sJ}^*(2860)$ to be  the $D_{s3}^*(2860)$ and $D_{s1}^*(2860)$ respectively, which differ from the experimental value $1.10 \pm 0.15 \pm 0.19$ greatly \cite{BaBar2009}. If the $D^*_{sJ}(2860)$ observed by the BaBar collaboration in the inclusive $e^+e^- \to \overline{D}^0K^{-}X$ production  and by the LHCb collaboration in the $pp \to \overline{D}^0K^{-}X$ processes  consists of  two resonances $D_{s1}^*(2860)$ and $D_{s3}^*(2860)$ \cite{BaBar2009,LHCb1207}, we expect to obtain an even smaller ratio $R$ in case of the   chiral symmetry breaking corrections  are small. On the other hand, if  the $D_{sJ}^*(2860)$ consists of  at least four resonances $D_{s1}^*(2860)$, $D_{s2}^*(2860)$, $D_{s2}^{*\prime}(2860)$,  $D_{s3}^*(2860)$, the large ratio $R=1.10 \pm 0.15 \pm 0.19$ is easy to account for, as the $J^{P}=2^{-}$
mesons $D_{s2}^*(2860)$ and $D_{s2}^{*\prime}(2860)$  only decay to the final states $D^*K$, see Eq.(9) and Eq.(15).

In the decays   $(1^-,2^-)_{\frac{3}{2}}\to (0^-,1^-)_{\frac{1}{2}}+ \mathcal{P}_j$, the ratio $R_{21}$
\begin{eqnarray}
R_{21}&=&\frac{\Gamma(2^- \to 1^-+\mathcal{P}_j)}{\Gamma(1^- \to 1^-+\mathcal{P}_j)} =  3 \, ,
\end{eqnarray}
 while in the decays    $(2^-,3^-)_{\frac{5}{2}}\to (0^-,1^-)_{\frac{1}{2}}+ \mathcal{P}_j$, the ratio $R_{23}$
\begin{eqnarray}
 R_{23}&=&\frac{\Gamma(2^- \to 1^-+\mathcal{P}_j)}{\Gamma(3^- \to 1^-+\mathcal{P}_j)} = \frac{ 7}{4} \, .
\end{eqnarray}
According to the ratios $R_{21}$ and $R_{23}$, the  $2^-$ state in a special doublet, irrespective of the $(1^-,2^-)_{\frac{3}{2}}$ doublet and the $(2^-,3^-)_{\frac{5}{2}}$ doublet,
 has a much larger  decay width to the final state $ 1^-+\mathcal{P}_j$ compared to its partner (the $1^- $ state or the $3^-$ state). The $2^-$ states in the $D_{sJ}^*(2860)$ can enhance the ratio $R$ significantly and account for the large ratio $R=1.10 \pm 0.15 \pm 0.19$ naturally. The ratios $R_{21}$ and $R_{23}$ are
 independent on the hadronic coupling constants and determined by the heavy quark symmetry and chiral symmetry. We can confront the present predictions to the experimental data in the future to examine the nature  of the  $D_{sJ}^*(2860)$ or identify the $D_{s2}^*(2860)$ and $D_{s2}^{*\prime}(2860)$ states. Furthermore, the chiral symmetry breaking
  Lagrangians ${\cal L}_{X}^\chi$ and ${\cal L}_{Y}^\chi$ have other phenomenological applications in the heavy-light meson systems, form example, we can study the strong decays of the D-wave $Q\bar{q}$ mesons and calculate  the scattering amplitudes of the S-wave and D-wave $Q\bar{q}$ mesons.

\section{Conclusion}
In this article, we take the $D_{s3}^*(2860)$ and $D_{s1}^*(2860)$ as the $1^3{\rm D}_3$ and $1^3{\rm D}_1$ $c\bar{s}$ states, respectively,  study their    strong decays  with the heavy meson  effective theory by including the chiral symmetry breaking corrections. We can reproduce the experimental value of   the ratio $R$, $R={\rm Br}\left(D_{sJ}^*(2860)\to D^*K\right)/{\rm Br}\left(D_{sJ}^*(2860)\to D K\right) =1.10 \pm 0.15 \pm 0.19$, with suitable hadronic coupling constants, the assignment of the  $D_{sJ}^*(2860)$ as the $D_{s3}^*(2860)$ is favored. The chiral symmetry breaking corrections are large, we should take them into account. Furthermore,   we obtain  the analytical expressions of the   decay widths, which can confronted with the experimental data in the future  from the LHCb,  CDF, D0 and KEK-B  collaborations   to fit the unknown coupling constants.
The present predictions of the ratios among the  decay widths can be used to study the decay properties of the $D_{s3}^*(2860)$ and $D_{s1}^*(2860)$ so as to identify them unambiguously. On the other hand, if the chiral symmetry breaking corrections are small, the large ratio  $R=1.10 \pm 0.15 \pm 0.19$ requires that the $D_{sJ}^*(2860)$ consists of at least four resonances $D_{s1}^*(2860)$, $D_{s2}^*(2860)$, $D_{s2}^{*\prime}(2860)$,  $D_{s3}^*(2860)$.

\section*{Acknowledgment}
This  work is supported by National Natural Science Foundation,
Grant Number 11375063,   and Natural Science Foundation of Hebei province, Grant Number A2014502017.

\end{document}